\documentclass[%
 reprint,
superscriptaddress,
 amsmath,amssymb,
rmp,
]{revtex4-2}
\usepackage[colorinlistoftodos]{todonotes}
\usepackage{xurl}
\usepackage{graphicx}
\usepackage{dcolumn}
\usepackage{bm}
\usepackage{graphicx}
\usepackage{soul}

 \usepackage[all]{background}
\usepackage{url}

\SetBgContents{\includegraphics[width=0.275in]{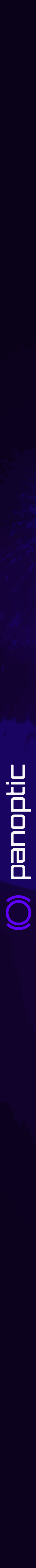}}
 \SetBgPosition{current page.west}
 \SetBgVshift{0.5cm}
 \SetBgHshift{0.4cm}
 \SetBgOpacity{1.0}
 \SetBgAngle{0.0}
 \SetBgScale{1.0}

\usepackage{tikz}

\begin{document}

\title{Panoptic: the perpetual, oracle-free options protocol}

\begin{tikzpicture}[overlay]
\node[minimum width=1in] (b) at (6.98,0.47){\includegraphics[width=2.4in]{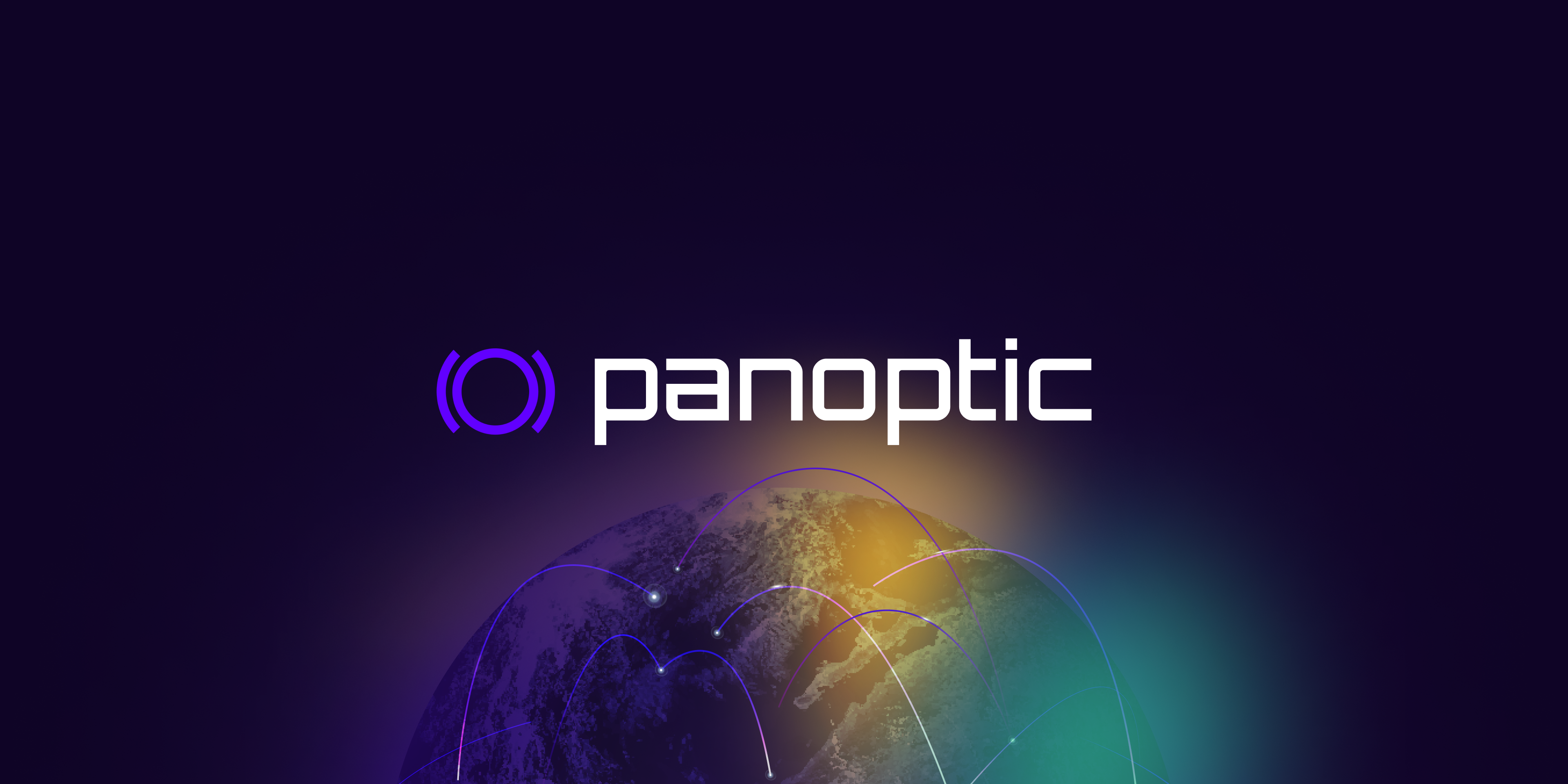}};
\end{tikzpicture}

\author{Guillaume Lambert}
  \email{glambert@panoptic.xyz}
\author{Jesper Kristensen}%
 \email{jkristensen@panoptic.xyz}

\date{June, 2023 v1.3.1}  


\begin{abstract}
Panoptic is the perpetual, oracle-free, instant-settlement options trading protocol on the Ethereum blockchain.
Panoptic enables the permissionless trading of options on top of any asset pool in the Uniswap v3 ecosystem and seeks to develop a trustless, permissionless, and composable options product, i.e., do for decentralized options markets what \texttt{x$\cdot$y=k} automated market maker protocols did for spot trading.
\end{abstract}

\maketitle

%

\section{Introduction}
\vspace{-0.5em}



Options are a foundational infrastructure for any healthy financial system, both in Traditional and Decentralized Finance. 
Options trading volumes for cryptographic assets like BTC and ETH have grown significantly in the last years~\cite{crypto_options} on centralized exchanges such as Deribit~\cite{deribit}, and on-chain options protocols have also recently gained significant traction.

According to DeFi Llama data~\cite{defi_llama}, there are already more than 30 on-chain options protocols with a Total Value Locked (TVL) exceeding \$830 million. 
While most of these options protocols emulate Traditional/Legacy Financial Markets (TradFi), the reliance on Call and Put options with fixed expiration dates and pre-specified strike prices fragments liquidity and makes adjusting positions expensive and cumbersome.
Moreover, accurate pricing of options via the \textit{Black-Scholes} model might not be readily achievable since all the operations that occur on-chain come with significant technical limitations.
Finally, on-chain option protocols are fundamentally limited by the quality and availability of their oracles, which rules out a large number in the long-tail of assets traded on decentralized exchanges. 

This whitepaper introduces the Panoptic protocol, a ``DeFi-native'' perpetual options protocol that aims to overcome the technical limitations and challenges that plague other on-chain options protocols.
Panoptic achieves this by relying on trading activity on Uniswap v3 \cite{uni3} to disintermediate the settlement of options contracts.
Specifically, the Panoptic protocol utilizes Uniswap v3 Liquidity Provider (LP) positions as a core primitive for trading long and short options \cite{guillaume_1}.
The Panoptic protocol consists of smart contracts that can directly interact with any Uniswap v3 pool contract to handle the minting, trading, and market-making of perpetual put and call options in a permissionless, trustless, and capital-efficient manner.

The Panoptic protocol implements the following key innovations:
\vspace{0.15em}
\begin{enumerate}
  \setlength\itemsep{0.25em}
    \item \textbf{Permissionless options}: All options in Panoptic are based on relocating liquidity within Uni v3 pools: moving liquidity closer to the spot price is a short option position, and moving it away from the spot price corresponds to a long option position.
    \item \textbf{Oracle-less Black-Scholes pricing}: Options pricing is based on a streaming premium model, where the premium initially starts at zero and increases over time according to a path-dependent pricing formula that converges to the Black-Scholes pricing model.
    \item \textbf{Distinct user roles}: Buyers and sellers will mint long and short options by directly interacting with the Panoptic smart contracts. Liquidity Providers (LPs) will provide fungible liquidity to the Panoptic pool without risking their capital.
    \item \textbf{Disentanglement of trading fees and LP rewards}:  LPs will accumulate yields by ``lending'' their liquidity to options sellers and buyers for a small fee. Options buyers and sellers will pay/receive a premium proportional to the fees generated by their Uni v3 options position.
    \item \textbf{Risk Management}: Options sellers will bear most of the risk in the protocol. Options sellers will be able to sell undercollateralized options (up to 5x capital efficiency). Multi-leg options instruments with up to four puts/calls can also be deployed in a single transaction.
    \item \textbf{Composability}: Multi-legged Panoptic options with user-defined payoffs can be tokenized as ERC1155 tokens. These tokens will be tradable and composable with other DeFi/staking protocols. 
\end{enumerate}
\vspace{0.15em}

The rest of the whitepaper covers i) Panoptic options (Panoptions) instrument based on Uni v3 LP positions; ii) an options premium model that accurately reflects the risk/reward of an option; iii) the Panoptic ecosystem and strategies for minting options; iv) how the protocol help its users mitigate risks; v) Panoptic's key design principles, applications and network effects.

\section{Panoptic Option Instruments}

Options in Panoptic\textemdash Panoptions\textemdash are distinct from TradFi options: they do not expire, they can be deployed at any strike, and their value does not decay over time. 
Panoptions trace their origin to the simple observation that providing concentrated liquidity in Uniswap v3 (Uni v3) is analogous to selling options in traditional finance \cite{guillaume_1}.
From the Uni v3 whitepaper~\cite{uni3}:
\begin{quote}
    Uniswap v3 is a noncustodial automated market maker implemented for the Ethereum Virtual Machine. In comparison to earlier versions of the protocol, Uniswap v3 provides increased capital efficiency and \textbf{fine-tuned control to liquidity providers}, improves the accuracy and convenience of the price oracle, and has a more flexible fee structure.
\end{quote}

Hence, unlike in Uni v2 where liquidity is distributed uniformly along the price curve from zero to infinity, concentrated liquidity allows LPs to allocate liquidity within a custom price range and on a set of price ticks (thus, in Uni v3 the price range is discrete). 
Concentrated liquidity thus improves capital efficiency and offers better control over where liquidity is deployed within a pool.
As outlined by Lambert \cite{guillaume_1}, Uni v3 liquidity positions effectively behave as two familiar types of options: \textit{cash-secured puts}\footnote{Check the glossary at the end of this whitepaper for definitions of all the terms used.} and \textit{covered calls} \cite{jesper_1}.

\begin{figure}
\centering
\includegraphics[width=3.5in]{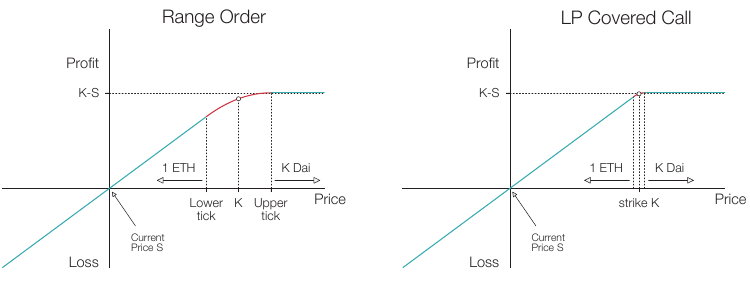}
\caption{\textbf{Left}: Deploying liquidity above the current spot price $\mathtt{S}$ (at origin) creates a range order which converts ETH to DAI as the price moves between the Lower and Upper ticks. The abscissa is the price of ETH vs Dai, the ordinate is the profit earned or the payoff curve, $\mathtt{K}$ is the price deployed to and serves as the strike price in the option-equivalent view. \textbf{Right}: Building on the left subplot, deploying liquidity to a narrow, single-tick range creates a ``covered call-like'' payoff where the ETH is sold immediately as the spot price crosses the strike $\mathtt{K}$.}
\label{fig:uni_lp}
\end{figure}

\begin{figure}
\centering
\includegraphics[width=3.5in]{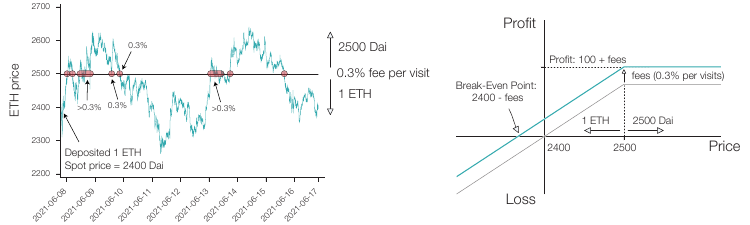}
\caption{LP covered calls. Single-tick liquidity accumulates a fixed 0.3\% fee each time the asset price visits the tick value.}
\label{fig:cc_fee}
\end{figure}

For instance, if we consider the value of a DAI-ETH LP position deployed to a single tick, the position will be 100\%~ETH exactly below the tick, and 100\%~DAI exactly above the tick, as illustrated in Fig.~\ref{fig:uni_lp}. 
The payoff of this position is identical to a covered call option.
Unlike regular options, however, LP tokens do not expire and assignment for Uni v3 options is reversible. 
Thus, the token composition of the LP position will shift between numéraire (e.g., DAI) and asset (e.g., ETH) every time the \emph{strike price} is crossed. 
Fees will be accumulated every time this crossing occurs (as shown for DAI-ETH in Fig.~\ref{fig:cc_fee}), and one could consider these fees as a short option premium that grows over time.
%
%


Hence, current liquidity providers in Uni v3 are already deploying \emph{short} option positions and are already receiving a premium-like compensation for the risk they take by collecting fees.
Theoretically, enabling the \emph{shorting} an LP position would effectively result in the same payoff as \emph{buying} a long option. 
However, implementing a lending protocol for Uni v3 LP positions is challenging because each LP position is unique and nonfungible.

The Panoptic protocol solves the constraints by facilitating the minting and lending of liquidity from Liquidity Providers to option sellers, making it possible to create long and short options based on tokenized concentrated liquidity positions in Uni v3.
The rest of this section provides more details about how Panoptic options work.

\subsection{Panoptic put options}
\begin{figure*}
\centering
\includegraphics[width=3.5in]{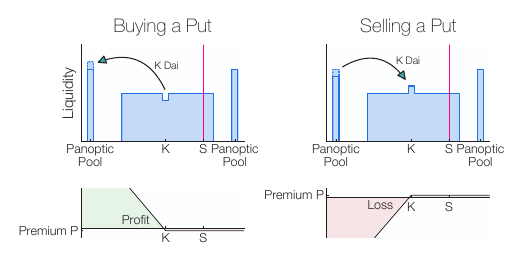}
\includegraphics[width=3.5in]{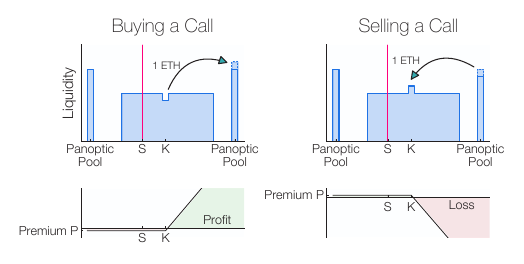}
\caption{Liquidity is moved away from the current price $\mathtt{S}$ for long (i.e., buying) options and moved closer to the current price for short (i.e., selling) options. In the top row, the x-axis represents the price of ETH in DAI and the y-axis is the liquidity amount deployed in each given price range. Note that the Panoptic pool represents a contract where funds are sent to and received from; in the illustration this is represented as a pool of liquidity far away from the price range. The bottom row shows the resulting payoff, or profit, curves. \textbf{Left}: Puts. \textbf{Right}: Calls.}
\label{fig:put_options}
\end{figure*}

Panoptic enables the deployment of long and short put options. 
The mechanism adopted by Panoptic for deploying a short put option at price $\mathtt{K}$ corresponds to locking $\mathtt{K}$ numéraire at a strike price $\mathtt{K}$ (Fig. \ref{fig:put_options}).
Conversely, removing $\mathtt{K}$ numéraire of liquidity at price $\mathtt{K}$ creates a long put option. 
In other words, users can create a long option by ``borrowing'' a short option. 

Let us consider what happens to an ETH-DAI option as the price crosses the strike price $\mathtt{K}$. 
When the spot price is above the strike price $\mathtt{K}$, the option is out-the-money (OTM). 
This means that the $\mathtt{K}$ DAI initially relocated to the $\mathtt{K}$ price tick is still worth $\mathtt{K}$ DAI. 
Thus the option may be closed by simply moving the $\mathtt{K}$ DAI back to the DAI liquidity pool with no cost. 

If the spot price is below the strike price $\mathtt{K}$, the option is in-the-money (ITM). 
If the position is a long put, closing the position would require the user to supply 1 unit of ETH to get $\mathtt{K}$ numéraire back, i.e., the put option guarantees that the buyer can sell 1 ETH for $\mathtt{K}$ DAI, regardless of the ETH price. 

If the position is a short put, the user would close the position by removing 1 ETH at strike price $\mathtt{K}$ and supplying $\mathtt{K}$ DAI back to the DAI liquidity pool, i.e., the option seller is obligated to purchase 1 ETH for $\mathtt{K}$ DAI, irrespective of the ETH price.

Note that an option writer/seller will relocate liquidity owned by the Panoptic liquidity pool, not their own when deploying an option. 
Similarly, the protocol will lock the relocated liquidity to the Panoptic liquidity pool when an option is purchased. 
This ensures that the Panoptic protocol can access the relocated funds and enables undercollateralized options writing (Section V).

\subsection{Panoptic call options}

In theory, long call options can be synthetically created by leveraging the \emph{put-call parity} and combining a long put with a long asset position. 
More details about the creation of synthetic call positions can be found in~\cite{guillaume_2}.
However, Panoptic avoids the creation of a short stock position because, at a fundamental level, a call at strike $\mathtt{K}$ in an ETH-DAI pool is identical to a put at strike $1/\mathtt{K}$ in a DAI-ETH pool (where the num\'eraire is now ETH).

In other words, when minting a short call, a user would need to remove 1/$\mathtt{K}$ ETH from the ETH liquidity pool and lock it at a \emph{strike price} $\mathtt{K}$. 
Similarly, a long call would remove 1/$\mathtt{K}$ ETH at a strike price $\mathtt{K}$ and lock it in the ETH liquidity pool. 

Fig.~\ref{fig:put_options} illustrates Panoptic mechanism for call options and the corresponding profit curve for long and short call options as a function of the spot price. 

If the spot price is above the strike price $\mathtt{K}$, the call option is in-the-money (ITM). 
If the position is a long call, the user would close the position by collecting $1/\mathtt{K}$ ETH from the ETH liquidity pool at depositing 1 DAI at strike price $\mathtt{K}$, i.e., a long call guarantees that the buyer can buy ETH for $\mathtt{K}$ DAI/ETH, irrespective of the ETH price.
Closing a short call would require the user to supply $1/\mathtt{K}$ ETH to the ETH liquidity pool and get 1 unit DAI back from strike $\mathtt{K}$, i.e., the short call owner is obligated to sell ETH for $\mathtt{K}$ DAI/ETH, irrespective of the price of ETH. 



\subsection{Limiting Gamma exposure}
\begin{figure}
\centering
\includegraphics[width=3.5in]{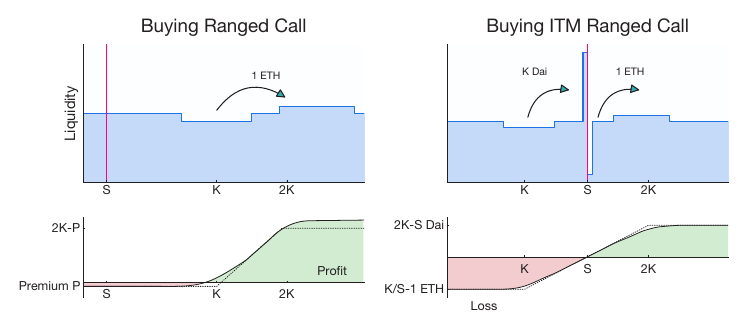}
\caption{\textbf{Left}: Long OTM call with wide range (fixed-gamma option) \textbf{Right}: Long ITM call with fixed-gamma.}\label{fig:fixed_gamma}
\end{figure}

One key advantage of Panoptic options over regular options is the ability to create options with a fixed range, resulting in an options with a fixed \textit{Gamma}. 
In other words, the \textit{Gamma} of an option can be capped by widening the range of an options position.
Examples of wide-range positions are shown in Fig. \ref{fig:fixed_gamma}. 

Specifically, options traders can utilize the relationship between the range factor $r={\sqrt{\textrm{priceUpper/priceLower}}}$ of a position and its effective days-to-expiration $T_r$ and an asset volatility $\sigma$ derived in \cite{guillaume3} and given by:

\begin{equation*}
T_{r} = \frac{2\pi}{\sigma^2}\left(\frac{\sqrt{r}-1}{\sqrt{r}+1}\right)^2.
\label{eq:eq0}
\end{equation*}

Controlling the range factor $r$ will effectively prevent options from ever reaching ``expiration'' and capping its \textit{Gamma} to $2/(K*\pi*\text{ln}(r))$.
Since $r>1$, the value of \textit{Gamma} will never diverge to infinity as the position approaches expiration. 
This will in turn completely eliminate pin risk, which manifests itself in TradFi when options approach expiration and rapidly shift between being ITM and OTM if the price is near the strike price. 

In addition, since pricing of Panoptions relies on a path-dependent mechanism (see Section III for more details), widening options also smooths out how quickly the price of an option increases over time.

\subsection{Panoptic ERC1155 composite option token}
Panoptic will allow users to sell undercollateralized options as a core element of the protocol (section IV).
This is in contrast to all other on-chain option protocols (except Opyn v2), which always require all positions to be 100\% backed by collateral.
Panoptic will allow the creation of undercollateralized options by using margin account collateral requirements similar to those developed by traditional financial institutions. 

\begin{figure}[b]
\centering
\includegraphics{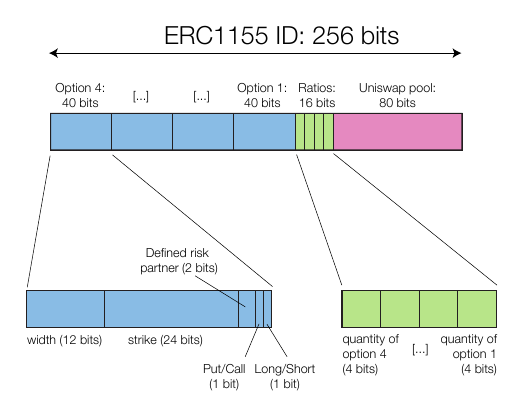}
\caption{Structure of a Panoptic composite ERC1155 option token. Up to four distinct options can be combined in a single ERC1155 token to facilitate collateral requirement tracking.}\label{fig:ERC1155}
\end{figure}

Another way Panoptic can reduce the collateral requirements of an option is by combining many of them into a single ERC1155 token to create \emph{defined-risk positions}. 
Specifically, the 256-bit ID of an ERC1155 will be used to encode information about up to four different options within the same pool.
The data encoded within each ERC1155 token is shown in Fig \ref{fig:ERC1155}. 

Combining several options into a single token allows the protocol to easily calculate the collateralization requirements of a set of interlinked options.
This is especially important when creating multi-legged options positions that may have a risk-defined profile even though each individual option may theoretically be exposed to infinite losses, e.g., a call spread's max loss is capped at the distance between the long and short call even though the short leg has infinity risk.

In addition, tokenizing complex option strategies could also facilitate the deployment of vault-like instruments which could, for instance, allow users to participate in vault based on short 16$\Delta$ strangles, 30$\Delta$ Jade Lizards, or synthetic equity positions.

\section{Oracles-less Black-Scholes Pricing}
Option positions in Panoptic have no expiration. 
This makes pricing them quite different from vanilla options whose price can easily be derived from the \textit{Black-Scholes} model given a time to expiration. 
While everlasting options with a constant funding rate have been proposed elsewhere~\cite{dave_white}, they rely on oracles and the constant rebalancing may be expensive to execute on-chain. 
To accurately price options, the Panoptic protocol will use a novel oracle-free concept called \textit{streaming premium}.

\subsection{Streaming premium model}

The  key difference between the pricing of regular options and the streaming premium model is that, instead of requiring the users to pay for their options upfront, the pricing of an option is path-dependent and will grow at each block according to the proximity of the spot price to the option strike price. 

Formally, this corresponds to continuously integrating the \textit{theta} of the option. 
Note that the theta of an option is defined as the derivative of the option's value with respect to time.
Assuming a zero risk-free interest rate, the theta of an option is defined as:

\begin{equation*}
\theta= \frac{\textrm{d}V(S,t)}{\textrm{d}t} = \frac{S\sigma}{\sqrt{8\pi t}}\textsf{exp}\left(-\frac{\left[\textrm{ln}\left\lbrace\frac{S_0}{K}\right\rbrace + \frac{\sigma^2t}{2}\right]^2}{2\sigma^2t}\right)
\label{eq:eq1}
\end{equation*}

Here, $S$ denotes the underlying asset spot (or current) price, $\sigma$ is the asset's volatility, $K$ is the strike price, and $t$ is the time to expiration. 
It is easy to show that one can recover the option's value by integrating theta over time and assuming that the price remains constant ($S_0$):

\begin{equation*} 
\begin{split}
V(S_0=\textrm{constant}, t) & = \int_{0}^{T} \theta (S_{0}, t)\textrm{d}t \\ 
&= \textrm{Call option price }C(S_0, K, T)
\label{eq:eq2}
\end{split}
\end{equation*}

If $S_0$ is not constant, however, one could ask whether it is possible to recover the call option price via integrating over the asset spot price history $S(t)$, i.e.:

\begin{equation}
\textrm{Premium P} = \int_{\mathcal{S}(t)} \theta (S_{t},K,\sigma)\textrm{d}t
\label{eq:eq3}
\end{equation}
This corresponds to integrating $\theta$ over the stochastic price path $\mathcal{S}(t)$. 

Computing the option premium using Eq. \ref{eq:eq3} may results in two unexpected extremal outcomes.
First, an option on an asset whose spot $\mathcal{S}(t)$ never comes inside the price range defined by range of the option will cost nothing. 
Second, if the asset price does visit the price range defined for the strike price, the option premium will increase according to the time it spends close (i.e., inside the price range) to the strike price, which may be several times larger that the Black-Scholes premium.

\subsection{Convergence to Black-Scholes pricing model}
Will ``integrating theta over the price $\mathcal{S}(t)$'' result in an fair options pricing? 
To answer this, we compute the average value of an option over all possible price paths $\mathcal{S}(t)$ by performing a \textit{Monte Carlo} simulation to compute the integral of theta over time for thousands of simulated price paths based on Geometric Brownian Motion.

\begin{figure}[t]
\centering
\includegraphics[width=3.6in]{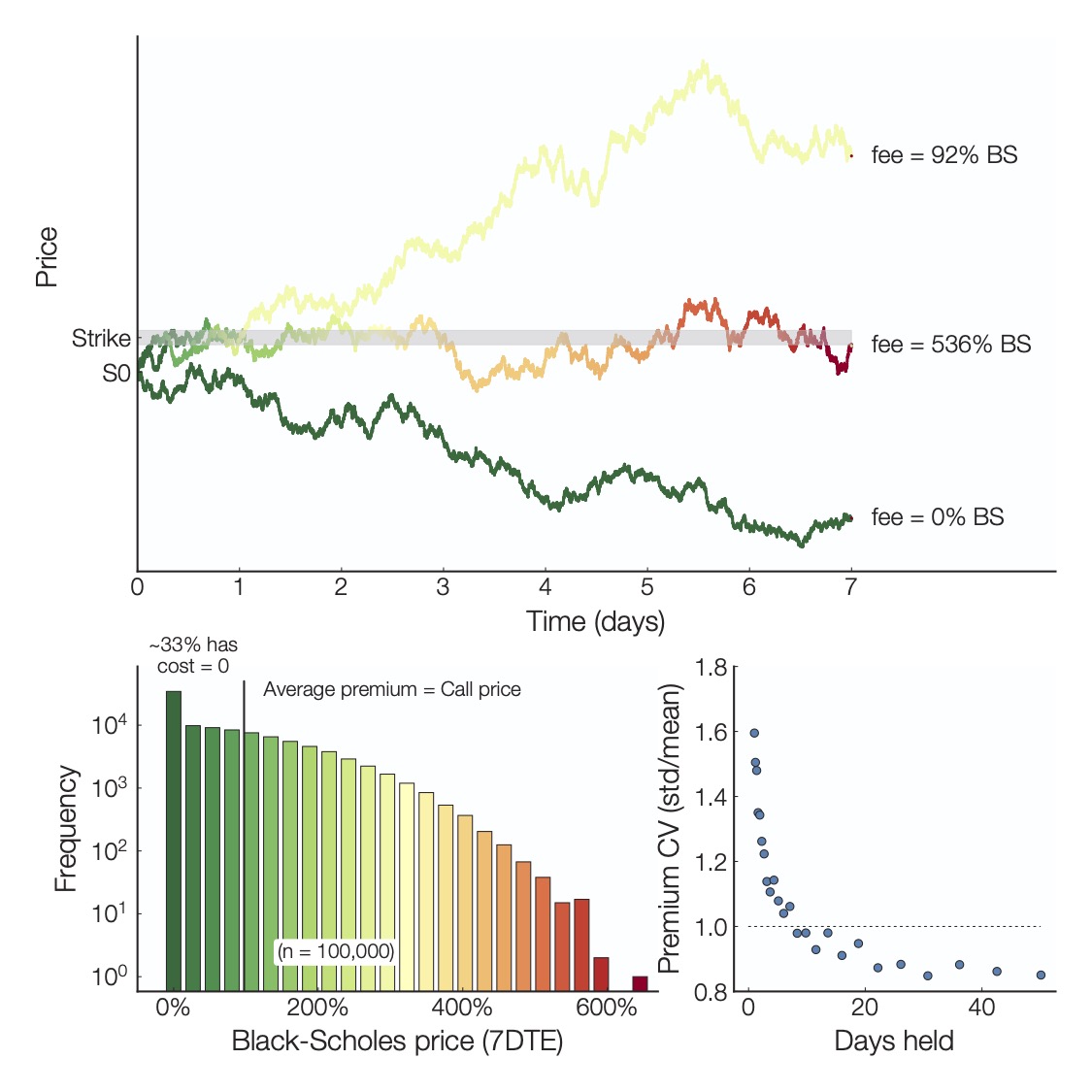}
\caption{\textbf{Top}: Three simulated price paths accumulate different amounts of fee/premium compared with the Black-Scholes price. \textbf{Bottom Right}: Distribution of the collected premium agrees with the Black-Scholes price on average. \textbf{Bottom Left}: Coefficient of variation (defined as standard deviation / mean) of the price of a streaming call option as a function of the time held. Parameters: Implied volatility=100\% per year taking a Monte Carlo step every minute.}
\label{fig:option_pricing}
\end{figure}

Interestingly, the result does converge to the \textit{Black-Scholes} price, but the distribution of option price can be quite large (Fig.~\ref{fig:option_pricing}). 
Indeed, since pricing depends on the specific price path, the option cost may be zero if the asset spot price never touches the strike price (i.e., never stays inside a price range defined by Panoptic), or it might be much higher in case the spot price touches the strike price repeatedly. 
We find that approximately 33\% of all streaming option premium for a call option held for 7 days would cost zero (Fig. \ref{fig:option_pricing}).


On the other hand, the actual cost of the option may be higher than Black-Scholes' prediction. 
Since users will only have to pay for an option if it is ITM, there is always a possibility that the asset spot price may touch the strike price many times and still end up OTM and have no intrinsic value. 
Our results show that 16\% of all streaming option premium would be twice as large as Black-Scholes.

Together, the simulation results indicate that the price of an option will heavily depend on the history of the asset price, with many options that spent all their time OTM being worth exactly zero or hovering around the strike price and being worth much more than the Black-Scholes price.
Fig.~\ref{fig:option_pricing} shows that the coefficient of variation, defined as the ratio of the standard deviation of the option price to its mean, approaches 82\% when held for more than ten days.


\begin{figure}[b]
\centering
\includegraphics{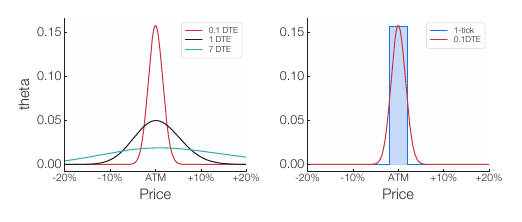}
\caption{\textbf{Left:} Theta vs. option price for different days to expiration (DTE). \textbf{Right:} 1-tick wide position as an approximation of a 0.1 DTE option's theta.}
\label{fig:theta_pricing}
\end{figure}

\subsection{Option pricing and implied volatility} 
As described above, the option premium is computed by directly integrating the price-dependent value of theta over time. 
One can think of a streaming premium pricing model as a series of continuously expiring options that accumulate a premium $\theta \Delta t$ at every time step $\Delta t$.

When it comes to choosing an appropriate value for $\Delta t$, let us first consult Fig.~\ref{fig:theta_pricing}, which shows the theta value of an option versus price, for 7 Days to Expiration (DTE), 1 DTE, and 0.1 DTE. 
As the time interval $\Delta t$ approaches zero, the option's theta starts looking like a Dirac delta function with a fixed width given by the full-width half max.

Hence, if we assume that the width of the delta function is the tick spacing of a Uni v3 pool (i.e., tickSpacing $tS$ = 0.02 for 1\%, 0.006 for 0.3\%, 0.001 for 0.05\%), then the value of theta can be approximated as the time spent inside the range multiplied by the height of the delta function. 
Since the area under the theta function is $k^2\sigma^{2}/2$ and the width is $k\cdot tS$ (prescribed by the tickSpacing $tS$), then the height will be $k{\sigma}^2/2tS$. 

This corresponds to a cumulative premium of $k^2\sigma^2/2tS \cdot \textrm{(time spent in range)}$. 
By comparing the cumulative premium to the actual fees collected by Uni v3 pool per unit time (i.e., $\mathtt{feeRate \cdot (Volume \cdot Time)/tickLiquidity}$), the implied volatility (IV), $\sigma$, can be derived as:

\begin{equation*}
\mathtt{IV \equiv \sigma= 2 \cdot feeRate \cdot  \sqrt{\frac{Volume}{tickLiquidity}}}
\label{eq:eq4}
\end{equation*}

Interestingly, this value for IV is in perfect agreement with the one derived using a completely different approach \cite{guillaume_4}. 
In other words, using the amount of fees collected as a measure of the option premium results in IV that depends on the traded volume and the amount of liquidity at the position's tick, and not on the option's realized volatility derived from actual market price fluctuations.

\section{Panoptic Ecosystem}
Panoptic ecosystem participants will include liquidity providers, options buyers, and options sellers. Their roles are summarized as follows:

\begin{description}
\item[Liquidity Providers] Provide fungible liquidity to the options market. They will be able to sell options against their liquidity at favorable rate (ie. zero commission), provided their collateralization ratio is larger than 100\%. 
\item[Option Sellers] Sell options by borrowing liquidity for a fixed commission fee (initially set at 0.1\%) and relocating it to a Uni v3 pool. Sellers have to deposit collateral and can sell options with notional values close to five times larger than their collateral balance.
\item[Option Buyers] Buy options by moving liquidity out of the Uni v3 pool back to the Panoptic smart contract for a fixed commission fee of 0.1\%. Buyers also have to deposit collateral (10\% of the notional value of the option) to cover the potential premium to be paid to the sellers.
\end{description}

\subsection{Liquidity Providers}
Liquidity Providers (LPs) will provide liquidity to the Panoptic smart contract by depositing assets into the option pool in the form of a single type (e.g., USDC or ETH). 
However, the net goal of LPs is to provide liquidity that can be borrowed and relocated to a Uni v3 pool rather than providing liquidity for spot trading directly.

At the time of deposit, the amount of liquidity deposited by the LP into the Panoptic pool will be recorded into the Panoptic smart contract and will be emitted as an ERC20 tokens tracking the LP's share of the pool. 

As buyers and sellers mint options, the liquidity deposited by the LPs in the pool will be deployed to the Uni v3 core pool contract. 
Option sellers will borrow the LPs' liquidity for a fixed commission fee (initially set to 0.1\% of the notional value) to create short put or call options.
LPs would anticipate that option sellers are savvy market participants that will optimize liquidity allocation to their benefits. 
Fees collected from that deployed position, which will be composed of both \texttt{token0} and \texttt{token1}, will be distributed to the option seller minus a \texttt{spread} that depends on the amount of available liquidity in the Uniswap pool. 

Option buyers will move liquidity from the Uniswap pool back to the Panoptic liquidity pools.
LPs will also collect a small commission based on the notional value of all relocated liquidity (initially set to 0.1\%).

LPs will also be able to sell options against their liquidity and they will pay zero commission fee as long as the value of the LP's collateral is larger than the notional value of all their options.
In other words, the commission will be waived if LPs sell options that are fully collateralized.


When a LP exits their liquidity from the option pool, Panoptic will burn their ERC20 tokens, and the LP will receive it share of the option pool, including collected fees. 
However, depending on the locked liquidity for open option positions, some of the LP liquidity might be unavailable.
That liquidity will be released/available once i) the long positions are closed; or ii) LPs force the exercise of far OTM long options.

\subsection{Option sellers}
Short options are minted by moving liquidity from Panoptic to the Uni v3 pool as long as it is below the spot price for puts and above the spot price for calls. 
Sellers will have to deposit some funds as collateral whenever selling puts and calls. 
The amount of collateral needed for deploying a short option will be less than 100\%: when an options is sold, the collateral requirement will be set to 20\% of the exercise value plus the option's ITM amount. 

The option seller will pay a fixed commission fee initially set to 0.1\% when minting an option that will be proportional to the notional value of the option (i.e., the amount of liquidity that was relocated).
Sellers may mint several option positions in a single transaction, and the combination of up to four options will be stored inside the smart contract as a single ERC1155 as an \textit{int256} bit. 
Combining options as a single ERC1155 instrument will greatly simplify collateral requirement computations.

When closing a short option, the user will have to repay the exact amount of tokens that was borrowed to the liquidity pool. 
Furthermore, they will receive an option premium from the liquidity pool that corresponds to the amount of fees that was collected by the deployed liquidity. 


\subsection{Option buyers}
Buyers may also mint several option positions in a single transaction, which will be stored inside the smart contract as a single \textit{int256} bit. 
Note that options can only be bought if option sellers/writers have already minted/sold them, hence make the options sellers an essential actor in the Panoptic ecosystem whose role may need to be incentivized.

The buyer will have to deposit some funds as collateral to cover future options premium to be paid to the option seller. 
The collateral amount equal to 10\% of notional corresponds to the maximum price that would be paid for a 45 DTE option with IV equal to 75\%.

The option buyer will also pay a fixed 0.1\% commission fee when minting option, which will be proportional to the notional value of the option (i.e., the amount of liquidity that was relocated).

When exercising a long option, the user will have to repay the exact amount of tokens that was borrowed to the liquidity pool. 
In addition, they will have to pay an option premium to the liquidity pool as a single asset, which corresponds to the amount of fees collected by the deployed liquidity plus a \texttt{spread} that depends on the liquidity in the Uniswap pool when the option was created. 
However, since ITM options involve the transfer of both tokens in a single transaction, the amount of premium paid will be deducted from the option's exercise amount (eg. deducted from the numéraire for puts and from the asset for calls)

\subsection{Total, Notional, and Locked Liquidity}

The amount of liquidity in a Panoptic pool will be tracked using three storage variables for the whole pool: $\mathtt{totalLiquidity}$, $\mathtt{totalNotionalValue}$, and $\mathtt{totalLockedLiquidity}$. 
The liquidity for each \emph{account} will also be tracked using three storage callable functions: $\mathtt{userLiquidity(userAddress)}$, $\mathtt{userNotionalValue(userAddress,\ positionList)}$, and $\mathtt{userLockedLiquidity(userAddress,\ positionList)}$.

\begin{itemize}
\item The $\mathtt{totalLiquidity}$ variable tracks the total amount of liquidity deposited by LPs.
The amount deposited by each user will be tracked using $\mathtt{userLiquidity(userAddress)}$.

\item The $\mathtt{totalNotionalValue}$ variable will track the amount of liquidity that has been moved to the Uni v3 pool, which is equal to the notional value of all the sold options.
The amount of notional value for each account will be reported using $\mathtt{userNotionalValue(userAddress, positionList}$, where the function requires the list of positions as an input.

\item The $\mathtt{totalLockedLiquidity}$ variable will track the amount of liquidity relocated from the Uni v3 pool to the Panoptic pool by option buyers. Liquidity needs to be ``locked'' because, when liquidity is moved back to the Panoptic pool by option buyers, it needs to be available when the option is exercised. 
The amount of locked liquidity for each user account will also be computed using the $\mathtt{userLockedLiquidity(userAddress, positionList}$ call function.
\end{itemize}
\emph{Note: since a user account may own hundreds of options positions, it is computationally difficult to manipulate an array of positions if it were stored in the smart contracts. It is easier to require the user to always supply the positionList as computed off-chain and have the smart contract loop through the provided list to check that 1) the user indeed owns each position, and 2) the number of positions matches optionBalance(userAddress).}

\subsection{Panoptic framework for cost calculation of options}

In order to calculate the premium/cost for the options, Panoptic will consider the total liquidity amount deployed at a given tick.
From the Uni v3 fee calculation formula, the fees collected are proportional to the amount of liquidity deployed. 
Assuming that the position is not in-range when fees are collected, the amount of fees for all positions deployed at a specific range is given by:

\begin{equation}
\begin{split}
\mathtt{totalFees = \left(fg_{upper} - fg_{lower} - fg_{insideLast}\right) \cdot liquidity}\nonumber
\end{split}
\end{equation}
where $\mathtt{fg_{upper}}$ and $\mathtt{fg_{lower}}$ are the $\mathtt{feeGrowthOutside0X128}$ of the upper and lower ticks, and $\mathtt{fg_{insideLast} = feeGrowthInside0LastX128}$, and $\mathtt{liquidity}$ is the liquidity owned by the protocol.
The option buyer would pay a premium equal to $\mathtt{totalFees \cdot positionSize / liquidity}$ to the seller.

One possible attack on the Panoptic protocol, however, would be for an options buyer to purchase all the options available and completely drain the liquidity at a specific tick.
In that case, the liquidity owned by the Panoptic protocol is zero (\texttt{liquidity = 0}), no fees will be collected (\texttt{totalFees = 0}), and the option premium will also be zero.

We solve this by tracking what we call an $\mathtt{effectiveLiquidity}$ for each position.
Specifically, instead of using \texttt{positionSize/liquidity}, the protocol will compute fees based on the actual amount of liquidity that has been deposited there by the option sellers \emph{before any option is purchased}.

For example, if 10 ETH worth of liquidity was sold at a given tick and a call option worth 9.9 ETH was purchased, only 0.1 ETH of liquidity remains in the Uniswap pool. 
However, the Panoptic smart contract base the premium calculation as if 10 ETH of liquidity was present even though only 0.1 ETH of liquidity is actually deployed. 

Using this argument, the cost of an option can be derived using an \texttt{effectiveLiquidity}, which means that the premium will be given by:
\begin{equation}
\begin{split}
\mathtt{Premium = totalFees}\cdot\underbrace{\mathtt{\frac{positionSize}{baseLiquidity - positionSize}}}_{\mathtt{effectiveLiquidity}}\nonumber
\end{split}
\end{equation}
where $\mathtt{baseLiquidity}$ is the amount of liquidity initially present before the options are created.
In the example above, the amount of fees paid will depend on $\mathtt{effectiveLiquidity} = 9.9/(10-9.9)$ = 99 ETH instead of \texttt{positionSize/liquidity} = 9.9/10 = 0.99 ETH.

The same argument can be made for short options (i.e., selling a new option should not consider the extra liquidity added to the pool at the time of deployment), meaning that the effective liquidity of a short option position is

\begin{equation*}
\begin{split}
\mathtt{effectiveLiquidity = \frac{positionSize}{baseLiquidity + positionSize}}\nonumber
\end{split}
\end{equation*}

The \texttt{effectiveLiquidity} should not differ much from the actual liquidity present when the \texttt{positionSize} is small relative to the amount of \texttt{baseLiquidity} owned by the Panoptic pool. 
This slight difference could be seen as a spread that is calculated for each option and increases for larger orders according to: $\mathtt{spread = } \pm~\mathtt{positionSize}/\mathtt{baseLiquidity}$.

Hence, Panoptic will store, for each option, a record of the $\mathtt{feeGrowthInside0LastX128}$ and $\mathtt{feeGrowthInside1LastX128}$ parameters at the upper and lower boundary of the LP position and the amount of $\mathtt{baseLiquidity}$ initially owned by LPs in the Uni v3 pool.

In addition, Panoptic will take special consideration for situations when a user purchases several small options rather than a large one. For instance, purchasing ten options each valued at 1 ETH should result in the same amount of premium as one option valued at 10 ETH. 

Therefore, if a user has already purchased $\mathtt{N1}$ options when the liquidity was $\mathtt{L1}$ and the user wants to deploy an additional $\mathtt{N2}$ options at a later time, then the \texttt{effectiveLiquidity} amount will be based on \texttt{baseLiquidity -= N2}, which results in and effective liquidity equal to $\mathtt{(N1+N2)/(L1-N1-N2)}$ for the combined options, and not $\mathtt{N1/(L1-N1)}$ for the first option and $\mathtt{N2/(L2-N1)}$ for the second one. 

\section{Risks}
\subsection{Collateralization requirements - option selling}

The selling of undercollateralized options is typically associated with significant risks for the Liquidity Providers: since they lend their liquidity to option sellers, they may lose their capital initially put forward if the user cannot cover the price of the option.

In traditional finance (TradFi), undercollateralization is handled by tuning the buying power requirement of an asset. For instance, the Cboe rules~\cite{cbos} state that the margin account collateral requirements for a short put or a short call are:

\begin{quote} 
    100\% of option proceeds plus 20\% of underlying security/index value less out-of-the-money amount, if any, to a minimum of option proceeds plus 10\% of underlying security/index value for calls; 10\% of the put exercise price for puts. (source: Cboe Rule 10.3)
\end{quote}
The Financial Industry Regulatory Authority (FINRA) Rule 4210~\cite{finra} also sets forth similar margin requirements for options.

In TradFi, if an OTM short put option at strike $\mathtt{K=\$49}$ is sold for a premium $\mathtt{P=\$1.50}$ while the current asset price is $\mathtt{S=\$50}$ and each contract corresponds to 100 units of stock, then the initial amount of collateral required to sell the option is

\begin{eqnarray*}
\mathtt{Collateral\ Requirement} &=& \mathtt{P}+0.2\cdot \mathtt{S} \cdot 100 - \mathtt{(K-S)} \\
&=& \mathtt{\$150 + \$1000 - \$100}\\
&=& \mathtt{\$1050}\\
\end{eqnarray*}

Therefore, the margin requirement would be approximately five times lower than the collateral requirement for a Non-Margin or Retirement account (collateral requirement = \$4900).

As a first step, the Panoptic protocol will employ requirements similar to the 20\% rule for most assets and instead compute the collateral requirements as:

\begin{quote}
    20\% of the \texttt{notional value} plus \texttt{max(ITM amount, 0)} minus \texttt{premiumAccrued}.
\end{quote}

Hence, a user wishing to sell a 2000 DAI-ETH put would initially need to provide \texttt{2000 * 0.2 = 400 DAI} of collateral.
This will be true regardless of the value of ETH because all puts can only be sold OTM (meaning that ITM amount is zero).
However, if the price were to move below the \texttt{2000} strike price, then the user would need to provide a ``maintenance'' margin equal to the ITM amount = $\mathtt{K-S}$.

As the protocol develops, the 20\% requirement may be a function of the pool utilization rate or the pool's implied volatility. 
Indeed, while the ``20\% of underlying security/index value'' is what is initially set by the Cboe and FINRA as a baseline level of collateral requirement for most securities, that requirement can and will be raised for highly volatile assets (up to 100\% in the case of GME in early 2021). 

There must therefore be a way to adjust the collateral requirements ``on-the-fly'' in response to a higher trading activity. 
One way to do this is by linking the collateralization ratio to the pool utilization (see section V.C below).

\subsection{Collateralization requirements - option buying}

Long options increase in price over time, so buyers will need to lock up collateral to ensure that they can still pay the option premium.

The collateral requirement for long options will also depend on the exercise value. 
However, since collateral requirement only needs to cover premium accrued, the collateralization ratio for long options will be 
\begin{quote}
    10\% of the \texttt{notional value} minus \texttt{max(ITM amount, 0)} plus \texttt{premiumAccrued}.
\end{quote}

\subsection{Pool utilization target}

Pool utilization is defined as the amount of liquidity in the Panoptic pool that has not been used to buy options. 
Specifically, 
\begin{eqnarray*}
\mathtt{Pool}&& \mathtt{Utilization} = \nonumber \\ \nonumber \\
&&\mathtt{\frac{totalLockedLiquidity}{totalLiquidity - totalNotionalValue}}.
\end{eqnarray*}

The optimal Pool utilization should be low enough to ensure that new options could be sold, but also high enough to reflect a healthy protocol use.

One way to decrease pool utilization (i.e., disincentivize option selling and buying) is to make the \ul{collateral requirement} a monotonically increasing function of the pool utilization rate. 

On the other hand, to increase the pool utilization (i.e., incentivize option selling and buying) is to make the \ul{commission} a monotonically decreasing function of the pool utilization rate.

Wherever the two functions meet would be the pool utilization target, where the commission and collateral requirement stabilized the pool utilization  towards an optimal level (with 50\% seeming like a reasonable target).



\section{Panoptic Protocol Advantages and Applications}
This section discusses the opportunities ushered by the Panoptic options protocols, focusing on  composability with other projects in the space. We also describe the several advantages Panoptic options have over regular options protocols, both in TradFi and DeFi on-chain.

\begin{figure*}
\centering
\includegraphics{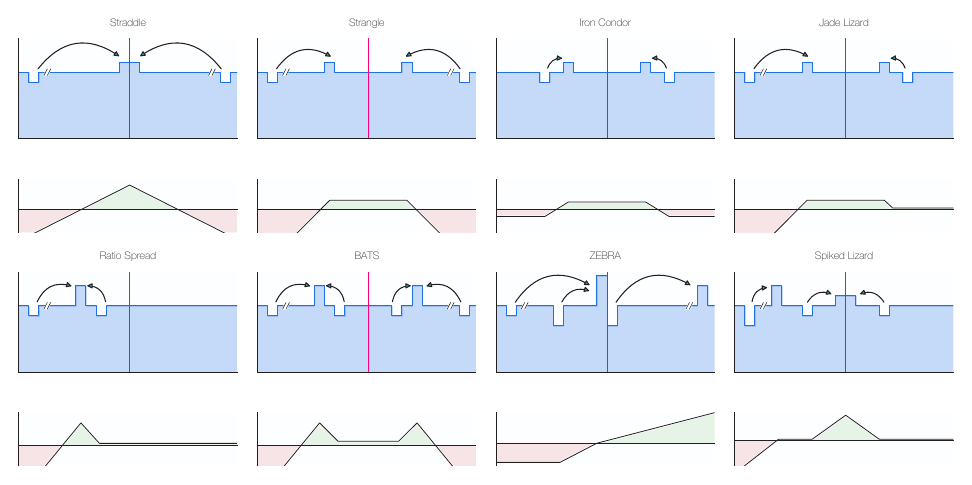}
\caption{Liquidity deployment and payoffs of composite multi-legged options that could be wrapped into a single ERC1155 token. \textbf{Top}:  Straddle (short ATM put+call), Strangle (short OTM put+Call), Iron Condor (OTM put+call debit spreads), Jade Lizard (short OTM put and OTM call credit spread). \textbf{Bottom}: Ratio spread (short put + put debit spread), BATS (put and call ratio spreads), ZEBRA (ATM short call and 2 ITM long calls), Spiked Lizard. }\label{fig:composite}
\end{figure*}







\subsection{Insurance protection for lending protocols}
Options can be utilized for protection/insurance against default by lending protocols. In other words, buying put options can substitute conventional liquidation frameworks used by the lending protocols and replace risk (i.e., liquidations failures) with a cost (i.e., buying put options).

\subsection{DAO treasury risk management}
Native tokens inside a DeFi project treasury might be financial resources. However, as explained in~\cite{hasu_1}, counting native tokens as assets on a balance sheet does much more harm than good for a DeFi project, and is considered poor treasury management. A more proactive and risk-averse approach for DAOs management would be to build treasuries with DeFi stable and blue chip assets and put these assets to work in relatively secure and established projects. At the same time, hedge the risk the DAOs have exposure to (by providing liquidity) via buying options.

\subsection{Tokenized risk positions}
Wrapped positions that make impermanent loss disappear: buy both puts to perfectly offset losses for Uni v3 position becoming OTM.

\subsection{Gamma-capped options}
Limiting \textit{Gamma} by using ranged positions: no pin risk, no Gamma explosion, only Theta (see Section II.c).

\subsection{Composite options}
Multi-leg tailored options with directional and/or strategic payoff structure can be minted via the composability feature of Panoptic. Fig.~\ref{fig:composite} highlights some of the composite options that can be created by relocating liquidity within the Panoptic pool accordingly.

\subsection{Capital efficiency}
The collateral deposited by the users never leaves the Panoptic smart contracts, enabling the system to remain solvent since each position is 100\% backed by collateral. Options sellers in Panoptic, however, will be able to write undercollateralized options that accurately reflect the risks associated with selling options. 

\subsection{Governance minimization}
Panoptic aims at limiting governance impact on how it is structured, i.e., no parameter should be based on governance decisions but rather function in a market-driven and responsive manner.

\subsection{Payoff compiler}
Panoptic introduced a novel concept called the $Gamma$ transformation \cite{guillaume_5}, which can compile any desirable payoff by composing a set of ticks. Each tick can be viewed as a ``Dirac delta" impulse, which translates to output being a given payoff. Like in, e.g., a Fourier transform, we can compose the desirable payoff via the correct composition of impulses.


\subsection{Greeks compiler}
In addition to the payoff compiler, we imagine the ability to construct any set of desirable Greek properties of the position. We imagine now a combined multi-objective optimization still posed as an inverse problem trying to find the set of impulses (the liquidity distribution) leading to a given joint payoff-greek distribution/behavior.


\subsection{Composability with on-chain option protocols}
The perpetual and cost-less options instrument offered by the Panoptic protocol may require seasoned options traders to reinvent themselves because most of the quantitative finance tools and metrics may not apply to non-expiring options. Therefore, other protocols may wish to build a protocol for expiring options on top of each Panoptic pool.
On the flip side, other protocols can tap into Panoptic to be able to build structured products such as options vaults.

\subsection{Hedging of off-chain options}
Panoptic options can be minted with a fixed gamma by specifying a width parameter. But since a fixed-width options is similar to an options with a perpetual Days to Expiration, a basket of Panoptic options could be used to continuously hedge off-chain options protocols. For instance, market participants that trades short options on a centralized exchange like Deribit could purchase a basket of fixed-width options on Panoptic that replicate a portfolio of expiring options without the need to rebalance over time.





\section*{Glossary}
\begin{itemize}
  \setlength\itemsep{0em}
    \item \textbf{Cash-secured puts}: selling a put option against cash the seller has set aside for buying the underlying asset.
    
    \item \textbf{Covered calls}: selling a call option against a long asset the seller already owns.
    
    \item \textbf{Credit call spread}: contain two calls (long and short) with the same expiration but different strike prices. The strike price of the short call is lower than the strike price of the long. The short call main purpose is to generate premium, whereas the long call helps limit the upside risk.

\item \textbf{Debit call spread}: contain two calls (long and short) with the same expiration but different strike prices. The strike price of the short call is higher than the strike price of the long call. The short call main purpose is to reduce the long call upfront cost.

\item \textbf{Delta}: measure of the change in an option price or premium resulting from a change in the underlying asset.

\item \textbf{Gamma}: measure of the rate of change in Delta over time, as well as the rate of change in the underlying asset. Gamma helps forecast price moves in the underlying asset.

\item \textbf{Implied Volatility (IV)}: the expected volatility of a stock/asset over the life of the option. In this whitepaper defined as $\sigma$.

\item \textbf{In-the-money (ITM)}: refer to an option that possesses intrinsic value. A call option is ITM if the underlying asset spot price is above the strike price. A put option is ITM if the underlying asset spot price is below the strike price.

\item \textbf{Out-the-money (OTM)}: refer to an option that only contains extrinsic value. A call option is OTM if the underlying asset spot price is below the strike price. A put option is OTM if the underlying asset spot price is above the strike price.

\item \textbf{Pin risk}: uncertainty that arises over whether an option's contract will be exercised (or assigned) when the expiration price of the underlying asset is at or very close to the option's strike price.

\item \textbf{Put-call parity}: require the call price plus the strike price of both options be equal to the underlying asset price plus the put price. When put-call parity is violated, an arbitrage opportunity exists.

\item \textbf{Put debit spread}: contain two puts (long and short) with the same expiration but different strike prices. The strike price of the long put is higher than the strike price of the short put. The short put's main purpose is to reduce the long put's upfront cost.

\item \textbf{Net delta}: the sum total of all the delta values of the options a trader owns. The closer the net delta position is to zero (delta neutral), the less it will be affected by changes in the price of the stock.

\item \textbf{Strike price}: price at which a put or a call option can be exercised.

\item \textbf{Theta}: measure an option's sensitivity to time. Consider an option with value 7.50 and a theta of 0.02. After one day, the option's value will be reduced to 7.48, 2 days to 7.46. etc. Theta loss is not constant like in this example but increases as the option approaches expiration.

\end{itemize}

\bibliographystyle{abbrvnat}
\bibliography{IEEEabrv,main}

\end{document}